YCr$_6$Ge$_6$ as a Candidate Compound for a Kagome Metal

Yui Ishii[1,2*], Hisatomo Harima[3], Yoshihiko Okamoto[1], Jun-ichi Yamaura[1], and Zenji Hiroi[1]

[1]*Institute for Solid State Physics, University of Tokyo, Kashiwa, Chiba 277-8581, Japan*
[2]*Department of Basic Science, University of Tokyo, Meguro, Tokyo 153-8902, Japan*
[3]*Department of Physics, Graduate School of Science, Kobe University, Kobe 657-8501, Japan*

We show that YCr$_6$Ge$_6$, comprising a kagome lattice made up of Cr atoms, is a plausible candidate compound for a kagome metal that is expected to exhibit anomalous phenomena such as flat-band ferromagnetism. Resistivity, magnetization, and heat capacity are measured on single crystals of YCr$_6$Ge$_6$, and band structure calculations are performed to investigate the electronic structure. Curie-Weiss-like behavior in magnetic susceptibility, $T^2$ dependence in resistivity, and a Sommerfeld coefficient doubly enhanced from a calculated value indicate a moderately strong electron correlation. Interestingly, the in-plane resistivity is twice as large as the interplane resistivity, which is contrary to the simple expectation from the layered structure. Band structure calculations demonstrate that there are partially flat bands slightly below the Fermi level near the Γ point, which is ascribed to Cr $3d_{3z^2-r^2}$ bands and may govern the properties of this compound.

[*] E-mail address: yishii@issp.u-tokyo.ac.jp

Itinerant electrons in a particular lattice such as a kagome or a pyrochlore lattice can form a unique band structure characterized by a flat band without dispersion. Various exotic phenomena associated with the flat band have been predicted in theoretical studies on the Hubbard model with nearest neighbor interactions.[1-3] For example, when the flat band is exactly located at the Fermi energy, the effective electron mass becomes infinitely large, resulting in an insulating state[4] in spite of the fact that wave functions are overlapped with each other so as to form a metallic state. Moreover, a flat-band ferromagnetism,[2,5] a fractional quantum Hall effect,[6] a heavy fermion state,[7-9] and exotic superconductivity[10] are expected when the flat band is half or partially filled. Despite these intriguing predictions, there have been very few experimental studies on the flat band, mostly because of the lack of suitable model compounds. In real compounds, deviations from the simple Hubbard model with the nearest neighbor interactions are inevitable. More importantly, a chance for a flat band to exist near the Fermi level may be rare, even if there is a flat band somewhere in the band structure. Therefore, it is a challenge for materials scientists to search for a suitable compound representing a kagome metal that exhibits exotic transport or magnetic properties arising from the flat band.

$Fe_3Sn_2$ is one of the compounds thus far studied as a candidate kagome metal compound. It crystallizes in a structure with a pair of distorted kagome lattices made of Fe atoms. A collinear ferromagnetic order is observed below 640 K, followed by a non-collinear ferromagnetic state, and finally, a spin-glass state appears below 80 K.[11] The relevance of this ferromagnetic order to the flat band ferromagnetism has not been clarified. In addition, an anomalous Hall effect is observed.[12] Competing exchange interactions between Fe spins may play a role in these complex magnetic properties. However, the presence of a flat band is not clear and its influence on the properties has not been explored. $CeIr_3Si_2$[13] and related compounds also have distorted kagome lattices consisting of transition metals, although there is no report focusing on the flat band.

We are interested in the family of compounds with the general formula $RT_6X_6$ ($R$ = Li, Mg, rare-earth elements, Zr, Hf; $T$ = Cr , Mn, Fe, Co, Ni; $X$ = Si, Ge, Sn).[14-16] Most of them crystallize in the $HfFe_6Ge_6$-type structure (space group $P6/mmm$) shown in Fig. 1.[14] This structure contains two undistorted kagome planes made up of the $T$ atoms in the unit cell, which are crystallographically equivalent and separated by two kinds of layer, the $RGe_2$ and $Ge_4$ layers. Their magnetic properties have been widely investigated thus far by focusing on the relationship between the localized *f* electrons of the rare-earth elements and the itinerant *d* electrons of the transition elements. Venturini *et al.* systematically studied the magnetic properties of a number of Mn and Fe compounds. They found that a ferromagnetic alignment of Mn/Fe spins often develops in the kagome layer and that various magnetic orders appear depending on interlayer couplings. For example, in $YMn_6Ge_6$ with a nonmagnetic *R* atom, Mn spins show a ferromagnetic order in the kagome plane, which are coupled antiferromagnetically perpendicular to the plane below 473 K, followed by a spin reorientation transition at 80 K.[17] A similar sequence of magnetic transitions has been reported in the Fe compound $YbFe_6Ge_6$.[18]

There are only a few studies on electrical resistivity and its anisotropy for $RT_6X_6$. In a $HoMn_6Ge_6$ single crystal, resistivity along the kagome plane is more than three times smaller than that perpendicular to the kagome plane,[19] as expected from the two-dimensional crystal structure. In fact, the corresponding anisotropy of the electronic structure is revealed in the band structure calculation on $HfMn_6Ge_6$.[20] However, the presence of a flat band arising from the kagome lattice is not observed in the band structure. Thus, the observed ferromagnetism of the compounds may not be related to the flat band ferromagnetism.

$YCr_6Ge_6$, which crystallizes in the $HfFe_6Ge_6$-type structure with an undistorted kagome lattice made of Cr atoms, shows a weaker magnetic response than the Mn or Fe compounds: a polycrystalline sample of $YCr_6Ge_6$ showed a weak temperature dependence of magnetic susceptibility with no magnetic order down to 1.8 K and a small ferromagnetic moment of 0.2 $\mu_B$ per Cr atom at the lowest temperature.[21] The electric resistivity and the band structure of this compound have not been investigated. In this study, we measure the magnetic susceptibility, heat capacity, and resistivity of $YCr_6Ge_6$ using single crystals and also investigate the electronic band structure via first principle calculations and the tight-binding approximation. We show that $YCr_6Ge_6$ is a potential candidate compound for studying the characteristics of the kagome metal.

Single crystals of $YCr_6Ge_6$ were grown by the flux method using a tin flux, following the method reported by Avila *et al.*[18] Y, Cr, and Ge powder and Sn shot (all with purity higher than 99.9%) were used as starting materials. A mixture of them at a ratio of Y : Cr : Ge : Sn = 1:3:6:20 or 1:1:6:20 was put into an alumina crucible and sealed in a silica tube with 0.2 atm Ar gas (99.9999%). After heating at 1000 °C for 3 h, the tube was slowly cooled to 500 °C at a rate of approximately 3 °C/h, and then furnace-cooled to room temperature. The product was immersed in hydrochloric acid in order to remove the tin flux, and then shiny single crystals were obtained. Two kinds of crystal with different morphologies were obtained depending on the starting composition: the 1:3:6:20 composition resulted in rod-like crystals with approximately 40 μm in diameter and approximately 2 mm in length, whereas the 1:1:6:20 composition gave hexagonal plate-like crystals with 500 μm edges. Typical crystals are shown in the inset of Fig. 2(b). Similar crystal growth with two types of morphology has been reported for $HoMn_6Ge_6$.[19]

Single-crystal X-ray diffraction experiments have revealed that the obtained crystals take the $HfFe_6Ge_6$-type structure. It is found that the long edge of the rod-like crystal is parallel to the $c$ axis, and the major surface of the plate-like crystal is perpendicular to the $c$ axis. The lattice parameters evaluated by means of powder X-ray diffraction on crushed single crystals are $a = 5.1688(3)$ Å and $c = 8.2793(5)$ Å, in good agreement with the previously reported values of $a = 5.167$ Å and $c = 8.273$ Å.[21] DC resistivity was measured by the four-probe method, and heat capacity was measured by the heat-relaxation method in a Physical Property Measurement System (Quantum Design). Magnetic susceptibility measurements were performed in a Magnetic Property Measurement System (Quantum Design).

Magnetic susceptibilities measured using approximately 10 plate-like crystals are shown in Fig. 2(a). $\chi_{//}$ and $\chi_{\perp}$ are magnetic susceptibilities measured in magnetic fields parallel and perpendicular to the $c$ axis. The former is approximately 20% larger than the latter at room temperature. This fact may reflect the two-dimensional character of the band structure of this compound. Upon cooling, both the $\chi$'s increase slightly, but there is no anomaly indicating a magnetic order down to 2 K. A saturated magnetization at 2 K was found to be as small as 0.002 $\mu_B$/f.u., which is much smaller than the reported value of 1.2 $\mu_B$/f.u. for polycrystalline samples.[21,22] Furthermore, this saturation magnetization did not depend on temperature up to room temperature. Thus, it may come from a small amount of ferromagnetic impurity having a high Curie temperature. On one hand, it is noticed that $\chi_{//}$ and $\chi_{\perp}$ show different temperature dependences at low temperature below ~20 K: the former becomes more enlarged than the latter as $T \to 0$. An additional Curie term from a magnetic impurity, which is often observed in low-dimensional or spin systems, should be isotropic and cannot explain this anisotropic behavior. It is noted that a similar behavior has been reported for a LuFe$_6$Ge$_6$ single crystal, although the origin is not clear.[18]

The absence of magnetic order or other phase transitions is shown in heat capacity measurements down to 0.4 K. The Sommerfeld coefficient $\gamma$ is estimated to be 66 mJ K$^{-2}$ mol$^{-1}$ from the $C/T$ vs $T^2$ plot shown in the inset of Fig. 2(b). According to our band structure calculations described below, the density of states (DOS) at $E_F$ is 200 states Ry$^{-1}$ (f.u.)$^{-1}$, which gives the band mass ($\gamma_{band}$) of 34.6 mJ K$^{-2}$ mol$^{-1}$. Thus, the observed $\gamma$ is enhanced almost twice as large as the $\gamma_{band}$. Provided that the room-temperature $\chi$ of 0.002 cm$^3$ mol$^{-1}$ represents the Pauli paramagnetic susceptibility $\chi_P$, the Wilson ratio $R_W$ becomes 2, indicating a strong electron correlation in this system.

Figure 3 shows the temperature dependence of the resistivity of YCr$_6$Ge$_6$. $\rho_{//}$ and $\rho_\perp$ represent resistivities measured in zero magnetic field with electric current running parallel and perpendicular to the $c$ axis, respectively. Rod-like and plate-like crystals were used for the two measurements, respectively. Both the $\rho$'s show metallic behavior, and there is no anomaly down to 2 K accompanied by superconductivity or other phase transitions. In the low-temperature range below ~15 K, the $T^2$ dependence is observed, which is characteristic for electron-electron scattering. However, the $T^2$ coefficients are small—$A = 4.7 \times 10^{-4}$ and $7.8 \times 10^{-4}$ μΩ cm K$^{-2}$ for $\rho_{//}$ and $\rho_\perp$, respectively—compared with those values expected for strongly correlated electron systems from the Kadowaki-Woods relation between $A$ and $\gamma$. This implies that the transport properties are governed by a light band rather than by the heavy band that is responsible for the mass enhancement observed in heat capacity.

Note in Fig. 3 that $\rho_\perp$ is larger than $\rho_{//}$; $\rho_\perp$ is almost twice as large as $\rho_{//}$ at 300 K. This is apparently in contradiction to the simple expectation from the two-dimensional crystal structure. In fact, in the related compound HoMn$_6$Ge$_6$, $\rho_\perp$ is approximately three times smaller than $\rho_{//}$ at 300 K.[19] This unexpected anisotropy in YCr$_6$Ge$_6$ must be attributed to the characteristic band structure of the compound.

We have investigated the energy band structure of YCr$_6$Ge$_6$ by means of first principle calculations by the full potential method based on the local density approximation and also by the tight-binding approximation. Figure 4(a) shows the former results. The structural parameters reported previously[22] were used in the calculations. It is worth noting that band manifolds with nearly flat dispersion appear slightly below $E_F$ near the Γ point toward the K and M points, which corresponds to the direction parallel to the layer in real space. In addition, another band manifolds with relatively flat dispersion exist above $E_F$ at approximately 0.69 Ry. On the other hand, there is a dispersive band from Γ to A, which corresponds to the direction along the $c$ axis in real space. The total and the partial DOS's are shown in Fig. 4 (b). Importantly, almost all the energy bands near $E_F$ are dominated by the Cr 3$d$ orbitals with negligibly small contributions from other orbitals. The small peak just below $E_F$ and the large peak above $E_F$ obviously correspond to the flat bands around the Γ point.

    From this characteristic band structure, one expects a considerably large mass enhancement for the in-plane flat bands and a relatively small mass for the dispersive band along the $c$ axis. Thus, the observed enhancement of $\gamma$ by a factor of two must be due to strong electron correlations associated with these partially flat bands of the Cr 3$d$ character just below $E_F$. Moreover, the observed anisotropy in the resistivity of $\rho_\perp > \rho_{//}$ can be understood by the presence of the light band along the $c$ axis; the in-plane resistivity tends to become large owing to the flat bands. The small $T^2$ coefficients may be ascribed to the light band along the $c$ axis, so that they are not enhanced owing to the electron correlations. Interestingly, one would expect a quasi-one-dimensional conduction along the $c$ axis if the flat bands were exactly located at $E_F$, in spite of the two-dimensional structural and electronic characters of the present compound. The anisotropy in the magnetic susceptibility of $\chi_{//} > \chi_\perp$ may also be due to the influence of the flat bands, although the mechanism is unclear.

    To get insight on the origin of the flat bands in YCr$_6$Ge$_6$, another band structure calculation has been performed on the basis of the tight-binding approximation using the Cr 3$d_{3z^2-r^2}$ orbital. Two-center overlap integrals $\sigma_n$ between $d$ orbitals of up to the fourth-nearest-neighbor Cr atoms are included in the calculations; as shown in Fig.1, the nearest- and fourth-nearest-neighbor atoms are within the same kagome plane, and the second- and third-nearest-neighbor atoms are in the adjacent kagome planes, which are separated by the Ge$_4$ and $R$Ge$_2$ intermediate layers, respectively. When only $\sigma_1$ is taken into account, a doubly degenerate flat band appears at $E = -2\sigma_1$ across the entire Brillouin zone, as well as two doubly degenerate dispersive bands. A small dispersion is induced

into the flat band when a relatively small $\sigma_4$ is switched on. On the other hand, the additional interplane interactions of $\sigma_2 = \sigma_3$ modify the flat band into a bonding flat band with a decreased energy along the Γ-K-M-Γ direction and an antibonding flat band with an increased energy along the A-H-L-A direction. As a result, a dispersive band appears along the Γ-A direction, *i.e.*, along the *c* axis. In addition, when $\sigma_2$ is not equal to $\sigma_3$, as in the case of YCr$_6$Ge$_6$, the degeneracy of the flat bands is lifted so that each flat band splits into two flat bands by energies $\sigma_2 + \sigma_3$ and $\sigma_2 - \sigma_3$ along the Γ-K-M-Γ and A-H-L-A directions, respectively.

In order to roughly reproduce the result of the first principle calculation shown in Fig. 4(a), the values of $\sigma_1$, $\sigma_2$, $\sigma_3$, and $\sigma_4$ are set to be -0.070, 0.010, -0.008, and -0.002 Ry, respectively. In addition, $\pi_n$ and $\delta_n$ integrals having magnitudes of $\pi_n = -0.5\sigma_n$ and $\delta_n = 0.1\sigma_n$ are also taken into account. Thus obtained tight-binding bands shown in Fig. 4(c) well reproduce the key features of the first principle calculation shown in Fig. 4(a). Now, it is clear that the partial flat bands originate from the Cr $3d_{3z^2-r^2}$ orbitals in the kagome lattice. The difference in energy between the Γ-K-M-Γ and A-H-L-A directions is caused by the interplane interactions.

The flat band observed in the present compound may not be a common feature in the family of $RT_6X_6$ compounds. In fact, such a flat band is not found at any energy in the band structure of HfMn$_6$Ge$_6$.[20] In addition, there is no such peak as in YCr$_6$Ge$_6$ in the calculated DOS of MgCo$_6$Ge$_6$.[23] One possible explanation is that $\sigma_4$ between Mn-Mn or Co-Co atoms in the kagome plane may be too large to destroy the flat dispersion. Of particular interest in YCr$_6$Ge$_6$ is that the lower flat band in the Γ-K-M-Γ direction happens to exist very close to $E_F$. This coincidence occurs because of the certain band filling and the appropriate magnitude of the interplane interactions. Therefore, the compound can be an exceptional candidate for the kagome metal.

Unfortunately, however, we have not observed dramatic phenomena in the present compound as are theoretically expected for the kagome metal. This is partly because the flat band is not completely flat and also because there are additional contributions from complex bands reflecting the actual crystal structure. A more serious problem is the fact that the flat band is not exactly at $E_F$ but located at 38 meV below $E_F$. According to our band structure calculations and assuming the rigid band picture, doping of approximately 0.5 hole per f.u. should reduce the Fermi level exactly to the flat band to make it emerge at the Fermi surface. Experimentally, this can be accomplished by the chemical substitution of Ga for Ge (Ga has one less electron than Ge). Looking for a drastic change originating from the flat band, experiments on systematic hole doping are now in progress.

In conclusion, YCr$_6$Ge$_6$ is a moderately correlated electron system having a kagome lattice made up of Cr atoms. It has a partially flat band in the band structure slightly below $E_F$, which originates from the Cr $3d_{3z^2-r^2}$ orbitals arranged in the kagome geometry. A mass enhancement by a factor of two is observed in heat capacity, and characteristic anisotropies are detected in resistivity and magnetic susceptibility, which are attributed to the influence of the flat band. YCr$_6$Ge$_6$ would potentially be a unique kagome metal when the Fermi level is finely tuned by hole doping so that the flat band emerges at the Fermi surface.

Acknowledgment
This work was supported by a Grant-in-Aid for JSPS Fellows.

Figure captions

Fig. 1. Crystal structure of YCr$_6$Ge$_6$. It crystallizes in the hexagonal HfFe$_6$Ge$_6$-type structure [$a$ = 5.1688(3) Å, $c$ = 8.2793(5) Å, space group $P6/mmm$], where two undistorted kagome planes made of Cr atoms are separated by Ge$_4$ and YGe$_2$ layers. The Wyckoff position of each site is indicated in parentheses. The overlap integrals $\sigma_n$ between the $d$ orbitals of Cr atoms up to the fourth-nearest neighbors are shown, and are used to calculate the band structure on the basis of the tight binding approximation shown in Fig. 4(c).

Fig. 2. (a) Magnetic susceptibility of plate-like single crystals of YCr$_6$Ge$_6$ measured upon heating in a magnetic field of 0.5 T. $\chi_{//}$ and $\chi_\perp$ represent those measured in magnetic fields parallel (open symbol) and perpendicular (solid symbol) to the $c$ axis, respectively. Inset shows a photograph of typical rod- and plate-like crystals on 1 mm grid paper. (b) Heat capacity measured on a plate-like crystal in zero magnetic field. Inset shows $T^2$ dependence of $C/T$ at low temperatures.

Fig. 3. Resistivities of YCr$_6$Ge$_6$ single crystals. $\rho_\perp$ and $\rho_{//}$ represent those measured with electric current perpendicular and parallel to the $c$ axis, respectively, and using plate-like and rod-like crystals, respectively. A $T^2$ plot is shown in the inset.

Fig. 4. Energy band structure of YCr$_6$Ge$_6$ near $E_F$ obtained by first principle calculations (a) and the corresponding densities of states (DOS) (b). Total DOS (black) and partial DOS from the Cr 3$d$ orbitals (dotted red) are shown in (b). Other contributions from the Cr $s$ orbitals, the Y $s$ and $d$ orbitals, and the Ge $p$ and $d$ orbitals are negligible. (c) Energy band structure obtained by the tight-binding approximation using the $3d_{3z^2-r^2}$ orbital on each Cr atom in the two kagome planes in the unit cell. Two-center overlap integrals $\sigma_n$ between $3d_{3z^2-r^2}$ orbitals up to the fourth-nearest-neighbor Cr atoms, which are shown in Fig. 1, are included in the calculations. The origin of energy is set to the energy of the $3d_{3z^2-r^2}$ orbital. In each figure, a couple of arrows indicate the positions of flat bands.

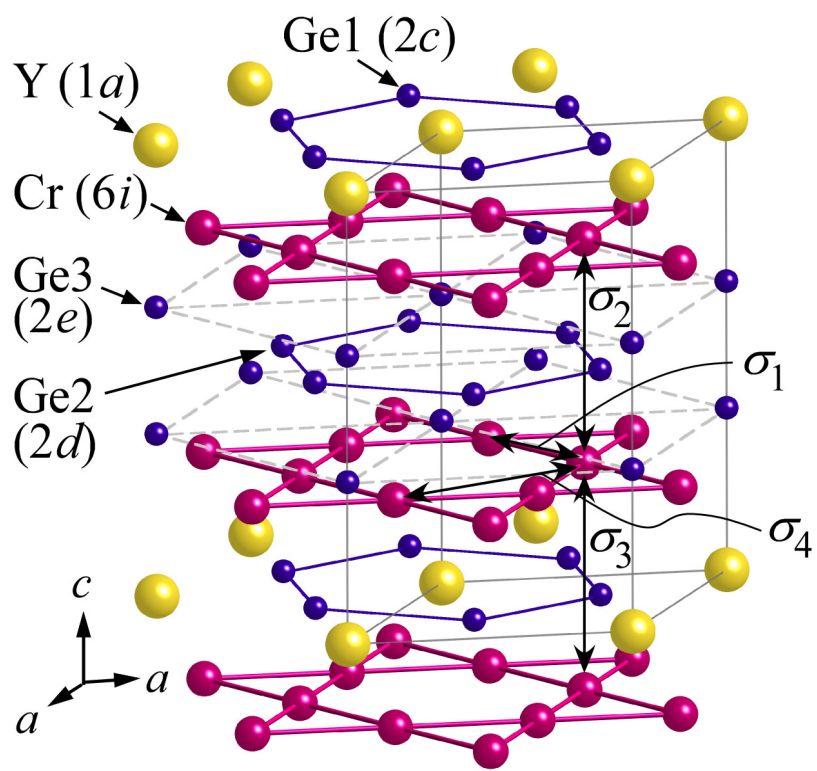

Ishii *et al.*, Fig. 1

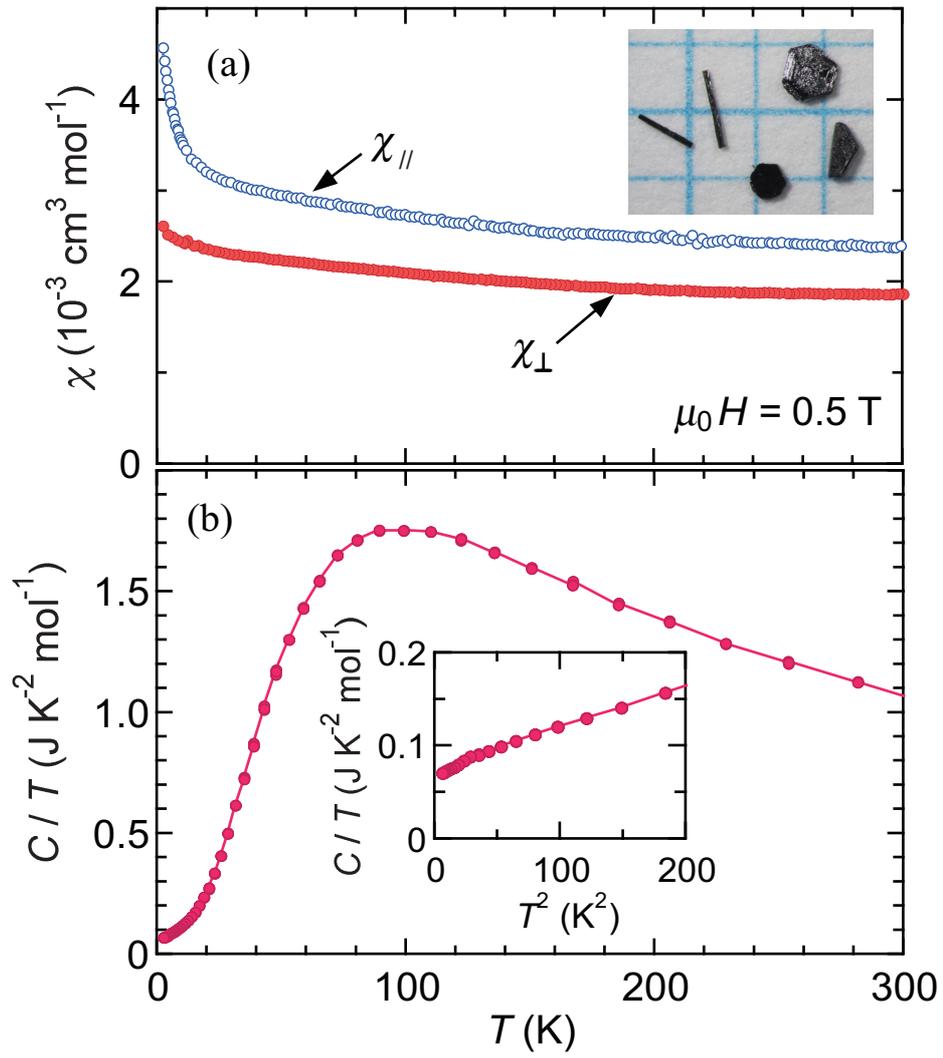

Ishii *et al.*, Fig. 2

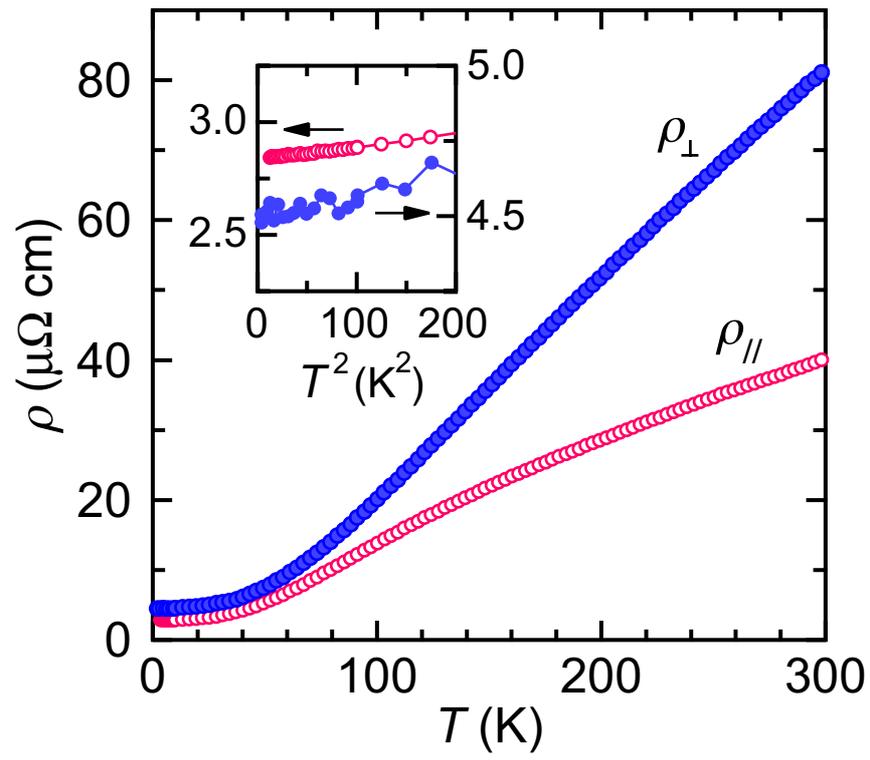

Ishii *et al.*, Fig. 3

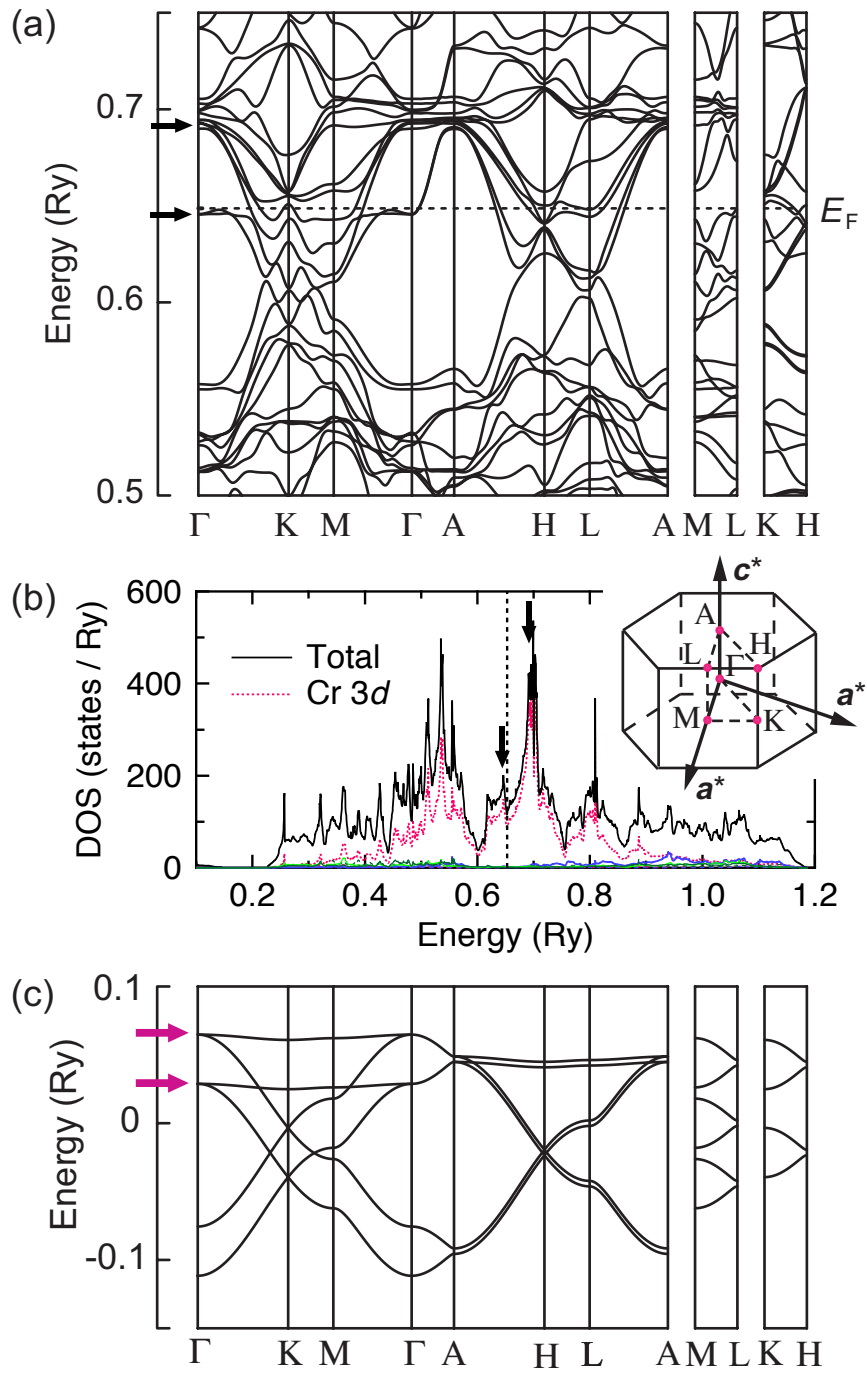

Ishii *et al.*, Fig. 4